\documentclass{iaus}
\usepackage{graphicx}

\title[secular Evolution and Morphological Transformation]
{Secular Evolution and the Morphological Transformation of
Cluster and Field Galaxies}

\author[Zhang and Buta]
{Xiaolei Zhang$^1$,
Ronald J. Buta$^2$}

\affiliation{$^1$Remote Sensing Division, U.S. Naval Research Lab,
\break 4555 Overlook Ave. SW, Washington, DC 20375, USA 
\break email: xiaolei.zhang@nrl.navy.mil\\[\affilskip]
$^2$Department Physics and Astronomy, University of
Alabama, \break 514 University Blvd E, Box 870324,
Tuscaloosa, AL 35401, USA \break email: buta@sarah.astr.ua.edu}

\pubyear{2006}
\volume{235} 
\pagerange{001--004}
\date{September 5th, 2006}
\jname{Galaxy Evolution across the Hubble Time}
\editors{F. Combes and J. Palous eds.}
\begin{document}

\maketitle

\begin{abstract}
Deep surveys conducted during the past decades have shown
that galaxies in the distant universe are generally of more
irregular shapes, and are disky in appearance and in their star
formation rate, compared to galaxies in similar environments
in the nearby universe. Given that the merger rate between
z=2 and the local universe is far from adequate to account
for this observed morphological transformation rate, an 
internal mechanism for the morphological transformation of 
galaxies is to be sought, whose operation can be further 
aided by environmental factors.  The secular evolution mechanism, 
especially with the discovery of a collisionless dissipation 
mechanism for stars within the secular evolution paradigm, 
has provided just such a framework for understanding the 
morphological evolution of galaxies across the Hubble time.
In this paper we will summarize the past theoretical results 
on the dynamical  mechanisms for secular evolution, and highlight
new results  in the analysis of the observational data, which
confirmed that density waves in physical galaxies possess the
kind of  characteristics which could produce theáobserved rates
of morphological transformation for both cluster and field galaxies.
\end{abstract}

\firstsection

\section{Introduction}

It has been a commonly-held belief that the stellar orbit in a galaxy 
containing quasi-stationary density
wave patterns does not exhibit secular decay or increase,
and there is no wave and disk-star interaction  
except at the wave-particle resonances (Lynden-Bell \& Kalnajs 1972). 
Zhang (1996, 1998, 1999) first demonstrated that secular orbital changes 
of stars across the {\em entire galaxy disk} are in fact possible due to a 
collective instability induced by the density wave patterns. 
The integral manifestation of this process 
is an azimuthal phase shift between the potential and density spirals, which 
results in a secular torque action between the wave pattern and the 
underlying disk matter. As a result,
the matter inside the corotation radius loses angular momentum 
to the wave secularly, and sinks inward. The wave carries the angular momentum 
it receives from the inner disk matter to the outer disk and deposits it 
there, causing the matter in the outer disk to drift further out.

\begin{figure}[ht]
\bigskip
\bigskip
\includegraphics[height=1.8in, width=1.8in,angle=0]{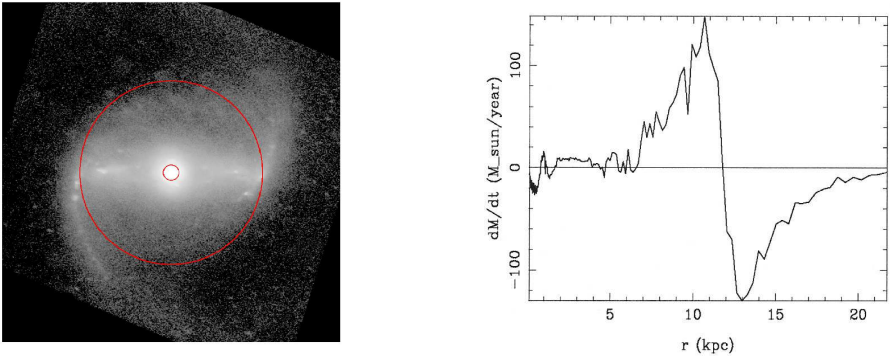}
\includegraphics[height=1.8in, width=1.8in,angle=0]{zhang_se_fig01_right.epsi}
\caption{{\it Left}: 
$K_s$-band image of NGC 1530 and the superimposed corotation circles
determined using the phase shift method (Zhang \& Buta 2006).
{\it Right}: Radial mass accretion/excretion rate calculated for
NGC 1530 from the $K_s$-band image.}
\end{figure}

The secular morphological evolution process causes the Hubble type of 
an average galaxy to evolve from late to early (Zhang 1999). 
The rate of secular evolution is expected to be faster for cluster galaxies 
than for isolated field galaxies because it is proportional to the wave 
amplitude squared and pattern pitch-angle squared (Zhang 1998), and 
cluster galaxies are found to have large amplitude and open density 
wave patterns excited through the tidal interactions with neighboring 
galaxies and with the cluster potential as a whole.  

\section{New Results on the Secular Mass Accretion Rates}

We have found in this study that physical galaxies generally contain
density waves strong enough to cause significant mass redistribution
over their lifetime.
In Figures 2, we show an image of the SBb galaxy NGC 1530 with superimposed
corotation circles, and the resulting secular mass accretion/excretion rate
calculated using the formulae given in Zhang (1998).
The calculated mass accretion/excretion rates for NGC 1530, 
with peak value around 145M$_\odot$/year, are
several orders of magnitude larger than for our Galaxy
($\sim$ 0.6M$_{\odot}$/year; Zhang 1999). This difference is 
due to the much larger density wave amplitude (average of 70\%),
larger pitch angle (65$^o$) and larger surface density (average of
100M$_{\odot}$/pc$^2$) of NGC 1530 compared to the Galaxy.
We have calculated the same quantities for several other galaxies and obtained
more moderate accretion rates, from a few to a few tens
of M$_{\odot}$/year.  In early
type galaxies the central mass accretion rate can be
much enhanced due to the presence of strong nuclear
density wave patterns, even though the outer wave patterns
generally become weaker as galaxies age.  Thus 
density-wave induced radial mass accretion 
could serve as the main driver for the morphological
transformation of both the field and cluster galaxies
over the past Hubble time, as well as for fueling active galactic
nuclei.  Furthermore, the mass accretion of the stellar
component (in addition to gas)
also means that most of the {\em stellar population} of the bulge
could form long before the {\em shape} of the bulge, maintaining
a homogeneous and older stellar population of the bulge during
the process of secular evolution.  This helps to solve one of the main 
difficulties in the secular building of bulges
from the accretion of the gas component alone, which leads
to a much varied stellar population in the resulting bulges.
The same phase-shift-induced mass redistribution process also
works in galaxies which contain other skewed density distributions
as observed in many high-z proto galaxies.

XZ acknowledges the support of the Office of Naval Research. 
RB acknowledges the support of NSF grant AST 050-7140 
to the University of Alabama.

\end{document}